\documentclass[aps,prd,showpacs,floats,nofootinbib,superscriptaddress,preprint,tightenlines]{revtex4}
\usepackage[mathscr]{eucal}
\usepackage[latin1]{inputenc}
\usepackage{amsmath}
\usepackage{amsfonts}
\usepackage{amssymb}
\usepackage{pictex}
\usepackage{epsfig}
\usepackage{colordvi}
\usepackage{color}

\def\bG{\bar{\Gamma}}

\def\be{\nopagebreak[3]\begin{equation}}
\def\ee{\end{equation}}
\def\ba{\nopagebreak[3]\begin{eqnarray}}
\def\ea{\end{eqnarray}}
\def\bas{\nopagebreak[3]\begin{eqnarray*}}
\def\eas{\end{eqnarray*}}

\def\d{{\rm d}}

\def\a{\alpha}
\newcommand{\teta}{\rlap{\lower2ex\hbox{$\,\tilde{}$}}\eta{}}

\newcommand{\bi}{\begin{itemize}}
\newcommand{\ei}{\end{itemize}}

\def\l{\lambda}
\def\g{\gamma}
\usepackage{enumerate}

\usepackage{colordvi}
\newcounter{mnotecount}[section]

\newcommand{\comment}[1]{}
\def\b{\beta}

\def\a{\alpha}

\def\t{\tau}
\def\k{\kappa}

\begin{document}
%\preprint{\vbox{\baselineskip=12pt \rightline{IGC-11/5-2}
%}}

\title{Effective constrained 
polymeric theories\\ and their continuum limit}
\author{Alejandro Corichi}\email{corichi@matmor.unam.mx}
\affiliation{Centro de Ciencias Matem\'aticas,
Universidad Nacional Aut\'onoma de
M\'exico, UNAM-Campus Morelia, A. Postal 61-3, Morelia, Michoac\'an 58090,
Mexico}
\affiliation{Center for Fundamental Theory, Institute for Gravitation \& the Cosmos,
Pennsylvania State University, University Park
PA 16802, USA}
\author{Tatjana Vuka\v sinac}
\email{tatjana@umich.mx}
 \affiliation{Facultad de
Ingenier\'\i a Civil, Universidad Michoacana de San Nicolas de
Hidalgo\\ Morelia, Michoac\'an 58000, Mexico}

\begin{abstract} The classical limit of polymer quantum theories yields a one parameter family of `effective' theories 
labeled by $\lambda$. Here we consider such families for constrained theories and pose the problem of taking the 
`continuum limit', $\l\to 0$. We put forward criteria for such question to be well posed, and propose a concrete 
strategy based in the definition of appropriately constructed Dirac observables. We analyze two models in detail, 
namely a constrained oscillator and a cosmological model arising from loop quantum cosmology. For both these models we 
show that the program can indeed be completed, provided one makes a particular choice of $\l$-dependent internal time 
with respect to which the dynamics is described and compared. We show that the limiting theories exist and discuss the 
corresponding limit. These results might shed some light in the problem of defining
the corresponding continuum limit for quantum constrained systems.
\end{abstract}

\maketitle

\section{Introduction}

A somewhat nonstandard and ``exotic" representation of the canonical commutation relations (CCR) in quantum mechanics 
\cite{exotic}, has received recently some attention in view of its possible relevance for some physical models 
\cite{AFW,jorma,CVZ2}. On its own, it represents an interesting model to study from the mathematical physics 
perspective. In particular, there has been some recent interest in exploring its relation with the standard {\it 
Schroedinger} representation of the CCR \cite{CVZ2}, a quantization that is privileged by the Stone-Von Neumann 
theorem. One feature that characterizes these new, {\it polymeric} representations is the appearance of a new 
dimension-full parameter $\lambda$ labeling them. In some cases this parameter can be thought of as providing a measure 
of discretization of space (or in some other cases, a lattice spacing). In all these cases, one expects to recover the 
standard theory in the limit $\lambda\to 0$ \cite{CVZ1,CVZ2,jorma}. In some of these instances, this limit has been 
referred to as a the {\it continuum limit} of the theory, borrowing the nomenclature used in lattice gauge theories. 
These new, polymeric representations are fundamental in the quantum treatment of simple cosmological models in what is 
known as {\it loop quantum cosmology} (LQC) \cite{lqc}. In that case, the corresponding parameter has the 
interpretation of providing a fundamental quanta for the spacial geometry. Thus, in LQC the limit $\lambda\to 0$ 
corresponds to the vanishing of `loopy'corrections to the theory, and one expect to recover then the standard 
Wheeler-De Witt quantum cosmology (see for instance \cite{slqc} and \cite{CVZ3}).\footnote{It should be noted that 
$\l$ captures the information about the non-local character of holonomies in loop quantum gravity. It is independent 
from $\hbar$ even when, in LQC, it carries the information about the discreteness of the spectrum of quantum geometric 
operators.}

Given that in some cases the parameter $\lambda$ can be seen as a `lattice regulator' whose continuum limit can be 
investigated, a natural avenue is to consider ideas from the renormalization group approach. The first steps were 
taken in \cite{CVZ1} for some simple quantum mechanical systems. An open issue is to even formulate the program when 
the system under consideration is subject to constraints. For instance, for totally constrained systems, there is no 
time evolution and one has to consider Dirac observables to describe the system. It is then not clear how to compare 
states and observables corresponding to different `scales' $\lambda$ and $\lambda'$, and even less how to approach the 
continuum limit. One certainly needs some new set of ideas to investigate this issue.

One important question that has emerged from the study of these quantum systems pertains to the existence of their 
semiclassical limit. That is, one might ask what is the classical theory that best describes the dynamics of 
semi-classical states. As it turns out, the corresponding classical description is {\it not} the classical system one 
started with, before quantization. Instead, one recovers a different classical theory where the `discreteness 
parameter' $\lambda$ plays a fundamental role. To be precise, one generically finds a corrected Hamiltonian 
$H_\lambda$ that controls the dynamics in such a way that, in the limit $\lambda\to 0$, one recovers both the 
classical Hamiltonian $H_0$ and the classical equations of motion. In what follows, we shall call these 
$\lambda$-dependent classical theories the {\it effective} theories, and the corresponding equations of the motion 
will be referred to as the effective equations.\footnote{In the LQC literature, where this nomenclature was first 
introduced \cite{SVV,APS}, the effective equations have been obtained using a rigorous procedure (see next footnote).} 
For instance, in the case of LQC the effective classical theory was analyzed in \cite{SVV,APS}, for a fixed value of 
$\l$, proportional to Planck length. However, it is perfectly valid and logically more generic to keep $\lambda$ as a 
free parameter and to explore changing it, while keeping the Planck length fixed (this has already been done at the 
quantum level in \cite{slqc} and \cite{CVZ3}). In this way we obtain a family of $\l$-dependent classical theories. 
Here we are interested in the corresponding classical ``continuum limit" $\l\to 0$ (with a slight abuse of notation).  
%We shall continue to use this nomenclature even when far from ideal.

Somewhat interestingly, in many cases one can obtain these effective equations by a simple substitution of some of the 
canonical variables by particular trigonometric functions (more details later). Thus, in this process, sometimes 
called {\it polymerization}, one starts with the classical Hamiltonian (or Hamiltonian constraint) $H_0$ and simply 
replaces some of the variables $q_i$ by $\lambda$-dependent periodic functions $f_i(\lambda)$ and arrives to the 
corresponding effective Hamiltonian $H_\lambda$.\footnote{This is not to be regarded as a fundamental derivation but 
rather as a shortcut to arrive to the effective equations. The correct procedure is to derive the effective equations 
from the polymer quantum theory by standard methods, that shall not be discussed here (See, for instance 
\cite{SVV,APS,victor,odceff,param-victor} for details). It should also be noted that there are examples in loop 
quantum cosmology, such as the $k$=1 model where the effective Hamiltonian can {\it not} be obtained by the simple 
polymerization method \cite{k=1}.} From this perspective, one could study the 1-parameter family of classical theories 
{\it per se} and consider their continuum limit, {\it with no reference to the corresponding quantum theory}. Thus, 
from the viewpoint of purely classical physics, one would be studying the family of $\lambda$-dependent theories and 
consider their $\lambda\to 0$ limit. This is precisely the problem that we shall consider here. The strategy will be 
simple: start with the family of effective theories and investigate the $\l\to 0$ limit.
 For simple, unconstrained systems this has been done before to some extent \cite{AFW,CVZ2,jorma}, but the same is not 
true of generic constrained systems.

The purpose of this article is to explore some of these issues. 
The first one is to put forward a formalism to deal with 
this family of classical constrained theories and to study what `taking the continuum limit' means. A short reflexion 
shows that the question is rather subtle. Unlike unconstrained systems, here each theory labeled by $\lambda$ lives on 
a different space, namely different constrained submanifolds $\bar\Gamma_\lambda \subset \Gamma$ of the common 
kinematical phase space $\Gamma$. Each physical configuration is an equivalence class of points on each of these 
submanifolds $\bar\Gamma_\lambda$, so how can we compare them? What is then the strategy to compare physical 
predictions from these theories and how does one define a limit? What is then the meaning of convergence of such 
limit? When can we say that the limit exists and that effective theories converge to another theory? Is this limit 
point equivalent to the original classical theory? In what follows we shall try to give answers to these questions. 
The second objective of this paper is to provide a conceptual framework that could be useful for the problem of 
answering the corresponding set of questions in the {\it quantum} theory. If we are able to answer some of these 
conceptual problems in the classical theory, this might help when constructing the corresponding continuum limit in 
the quantum realm. We shall not attempt to do that in this manuscript \cite{CVZ-Future}.

\begin{figure}
 \centering
  \epsfig{file=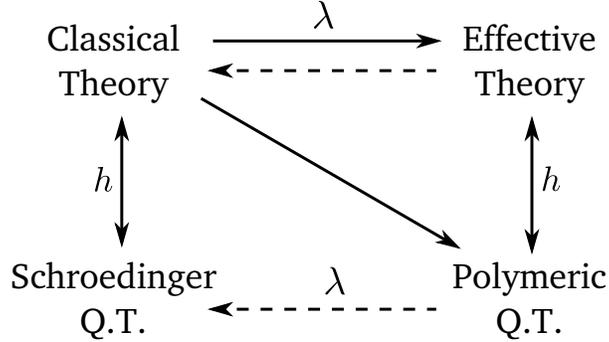,scale=1.4}
\caption{Connection between theories on various levels: classical, classical-quantum and quantum.}
\label{fig:1}
\end{figure}

The relation between all these theories is illustrated in Fig.~\ref{fig:1}. The vertical arrows correspond to changing 
Planck's constant $\hbar$, with the top row corresponding to $\hbar$=0. The horizontal arrows depicts changing the 
value of $\l$, with the left column corresponding to $\l$=0. Thus, the top left corner is the classical theory. 
Following the vertical arrow down corresponds to the standard process of Dirac-Schroedinger quantization. The vertical 
arrows in the upward direction correspond to the ``classical limit". The diagonal arrow is the so-called polymer 
quantization, that starts with the classical theory and provides a polymer quantum theory. The left pointing arrow in 
the bottom row, corresponds to the so called ``continuum limit" that starts from the $\l\neq 0$ polymer quantum theory 
and defines the limit $\l\to 0$. The process called `polymerization' corresponds to the right pointing arrow in the 
top row. The problem we are considering here, namely the continuum limit for classical theories, corresponds to the 
left pointing arrow in the top row that, as we shall here argue, does not lead in general to the original classical 
theory. This fact is depicted in the figure as a dotted arrow. Note that we shall {\it not} discuss in this note the 
quantum theories, nor their semiclassical and continuum limits.

The strategy we shall follow can be stated rather succinctly. We shall define and construct appropriate physical 
observables that will serve for the purpose of comparing theories at different scales. One possible approach would be 
to isolate the corresponding reduced phase spaces $\hat\Gamma_\lambda$ and to define a mapping between them such that 
observables are `pullbacked' and compared. While this strategy seems natural, it can only give us insight into the 
{\it frozen} description of dynamics. How can we then consider the `dynamical' evolution --even when generated by 
constraints-- in this formalism? For that we shall follow the strategy put forward by Rovelli, Dittrich and others to 
construct relational (Dirac) observers that describe dynamics even in the totally constrained case 
\cite{rovelli,dittrich}. But the choice of such observables and thus of the notion of dynamics is far from being 
unique. Does this choice has then an impact on the notion of convergence and the limit? As we shall see in detail, the 
answer is in the affirmative. It is then a non-trivial task to select the correct observable that shall play the role 
of time, for which the reconstructed dynamics converges. Taking the limit then means taking the corresponding 
observables and comparing them in the limit with, for instance, the alleged `continuum theory'. Using this as our 
guiding principle we shall see that in two examples of interest, one can indeed find such $\lambda$-dependent time 
observable, with respect to which the question of convergence can be posed. As we shall see clearly illustrated by 
these examples, this choice is unique, which implies that the notion of convergence and the existence of a `continuum 
limit' has some limited domain of validity. That is, convergence can not be generically realized for arbitrary 
observables and time functions.

The structure of the paper is as follows. Section~\ref{sec:2} corresponds to the core of the paper. In this section, 
we define the class of theories we are interested in, construct the corresponding effective theories and discuss the 
notion of convergence we shall consider. We end by specifying the type of observables that one should construct for 
such purposes. In Section~\ref{sec:3} we give our first example, namely a parametrized harmonic oscillator and 
corresponding effective theories. We construct Dirac observables and analyze their convergence. Section~\ref{sec:4} is 
dedicated to the convergence of effective classical theories based on a solvable model of LQC. In Section~\ref{sec:5} 
we summarize our results and give some ideas for future work.

\section{Formulation of the problem}
\label{sec:2}

As we have discussed in the Introduction, our objective is to consider one parameter families of `polymeric
classical theories' in totally constrained systems and to study its possible limit --and convergence-- when taking 
the limit $\lambda\to 0$. In this section we shall be more precise about the different steps that need to be 
taken in order to achieve this. This section has three parts. In the first one, we shall specify the type of constrained theories under consideration. 
In the second part, we describe in detail our strategy to identify configurations at different 
scales in order to be able to take the limit and we specify our notion of convergence.
In the third part we shall introduce complete Dirac observables constructed from the so-called 
partial observables. 

\subsection{Class of theories}

%In this work we are concerned with convergence of the appropriately constructed effective gauge theories.
We start with a finite dimensional classical theory that describes a system with one constraint ${\cal C}$ on 
the kinematical phase space $\Gamma$. The elements of $\Gamma$ where the constraint vanishes represent points on the
`classical' constraint surface $\bar{\Gamma}$. The constraint generates canonical transformations on $\bG$ that, 
in the standard interpretation, are regarded as gauge transformations.
Let us now specify the family of effective gauge theories, characterized by the parameter $\l$, with
constraint ${\cal C}_\l$.

Suppose that the finite dimensional kinematical phase space $\Gamma$ is parametrized by canonical coordinates $(q_i,p_i)$, $i=1,\ldots ,n$, with $\{q_i,p_j\}=\delta_{ij}$. 
The classical theory is defined by the constraint
\be
 {\cal C}(q_i,p_i)\approx 0\, .
\ee
A special case which is of interest to us is when the constraint is of the following form,
\be
 {\cal C}(q_i,p_i)\equiv {\tilde{\cal C}}(q_1,\dots q_{n-1},p_1,\dots p_{n-1})+\a (p_n)^k \approx 0\, ,
\ee
where $\a$ is an arbitrary parameter and $k>0$. Now, since the constraint does not depend on $q_n$, its canonically conjugate momentum $p_n$ is a Dirac observable ($\{p_n,{\cal C}\}=0$), so it remains constant along the
gauge orbits, while $q_n$ is a linear function of the `affine parameter' $t$ along the orbit, and can
be viewed as an evolution parameter. 

Now we shall construct a one parameter family of effective theories. 
This particular construction is motivated by the results from polymer quantization,
where, in one dimension, the derivative operator that would correspond to the operator $\hat{p}$ does not exist, but the Weyl operator corresponding to finite translations along $q$ {\it is} well
defined. In this case we can approximate the operator acting on state $f(q)$ corresponding to the derivative with respect to $q$ as
\ba
\partial_q f(q) & \approx & \frac{1}{2\l}\bigl[ f(q+\l )-f(q-\l )\bigr] \nonumber\\ &=& \frac{1}{2\l}(\widehat{e^{ip\l}}-\widehat{e^{-ip\l}})f(q)
=\frac{i}{\l}\;\widehat{\sin(\l p)}\;f(q)\, .
\ea
Similarly, the second derivative can be approximated by
\be
\frac{1}{\l^2}\bigl[ f(q+\l )-2f(q)+f(q-\l)\bigr] = \frac{2}{\l^2}\,\bigl( \widehat{\cos{\l p}}-1\bigr)\, f(q)\, .
\ee
where $\l$ is the free parameter that represents the displacement along $q$ in the approximation. Note that by keeping $\l$ finite, the process sometimes called `polymerization' could be defined as the substitution
\be
\hat{p} \longrightarrow \frac{1}{\l}\,\widehat{\sin(\l p)}\qquad {\rm and}\qquad 
\hat{p}^2 \longrightarrow \frac{2}{\l^2}\,(\widehat{\cos(\l p)} -1)\, ,\label{subs}
\ee
in the Hamiltonian or any other operator of the theory.
This substitution, that is defined in the quantum theory, suggests that one can also define a
`polymerization' of the classical theory. Let us now see how that comes about. 
Consider a function $F(p,q)$ on phase space. One can then define its {\it polymer transform} ${\cal P}[F]$
as the function $F$ with the substitution given by (\ref{subs}). Thus, we have
\ba
{\cal P}[F(q)]= F(q)\qquad &;&\qquad {\cal P}[p]=\frac{1}{\l}\,\sin(\l p)\\
{\cal P}[p^2] &=&
\frac{2}{\l^2}\,(\cos(\l p) -1)
\ea
and similarly for higher powers of $p$.\footnote{It has sometimes been suggested that by polymerization of $p$, one
could consider other periodic functions that behave as $p$ when $\lambda p\ll 1$, such as $\tan(\l p)/\l$. However, the choice of $\sin(\l p)/\l$ is uniquely selected since it comes, as we have seen, from the discretization of the
derivative in the quantum theory.}

We are interested in the analysis of the effective classical theories defined by the `effective constraint'
\be
{\cal C}_{\l}(q_1,\dots q_{n-1},{\cal P}[p_1],{\cal P}[p_2],\dots {\cal P}[p_{n-1}], p_n)\approx 0 \, . 
\ee
%Where ${\cal P}[p_i]$ is the polymeric transformation of $p_i$, meaning that in the original classical
%theory we should replace $p_i$ with $\frac{1}{\l}\sin{\l p_i}$, $p_i^2$ with $\frac{2}{\l^2}(\cos{\l p_i}-1)$,
%and analogous expressions for higher power of $p_i$. 
Note that the moment we can consider ${\cal P}[p_i]$ instead of $p_i$, that coordinate becomes effectively compactified, given that the Hamiltonian (and all observables) are now periodic functions of $p_i$. One could still regard $p_i$ as unbounded, but with the restriction that all physically relevant observables are periodic. 

It is clear that we have a one parameter family of ``effective Hamiltonian constraints" ${\cal C}_\l$ with the
property that
\be
\lim_{\l\to 0}{\cal C}_\l ={\cal C}\, .
\ee
Each of these constraints defines a submanifold of $\Gamma$, the corresponding $\l$-dependent constrained surface
$\bar{\Gamma}_\l\subset \Gamma$ defined as those points of $\Gamma$ where ${\cal C}_\l$ vanishes. Clearly, the
submanifolds $\bar{\Gamma}_\l$ do not intersect for $\lambda\neq\lambda'$, so what we have is a one parameter
family of distinct constrained theories that foliate a region of $\Gamma$.

Let us now pose the question of convergence,
with the goal of comparing physical configurations on each of these submanifolds.

\subsection{Convergence}

The first challenge in our program to define a `$\l$-flow' in our space of theories (each one labeled by $\l$), is to provide a prescription for identifying physical configurations at different scales. Since we are dealing with totally constrained systems, physical states correspond to gauge orbits under the flow generated by the constraint ${\cal C}_\l$. In other words, they correspond to equivalence classes under the constraint flow or, equivalently, points in the corresponding reduced phase space $\hat{\Gamma}_\l$. How are we to define a mapping ${\cal R}_{\lambda,\lambda'}:
\hat{\Gamma}_{\l'}\rightarrow  \hat{\Gamma}_{\l}$? Are there any preferred criteria motivated by physical or consistency considerations? It is clear that among such criteria, one should ask that, for two scales $\lambda,\lambda'$ that are close to each other (say, satisfying $|\lambda-\lambda'|<\epsilon$), the two gauge orbits should be `close to each other'. But, what does `close to each other' mean when they lie on different spaces? 
How can we even hope to compare them?

In order to define such correspondence we need to analyze the behavior of the gauge orbits for different values of $\l$. For instance, we know that the `equations of motion' for all theories are very close to each other when
$\l p\ll 1$.  Thus, one would expect that the orbits follow very similar trajectories in a neighborhood of such plane. Thus the strategy to the find the mapping ${\cal R}$ is to identify, in that neighborhood, orbits that are close to each other (as seen in the large kinematical space $\Gamma$), by fixing some constants of the motion, such as $p_n$ and other `initial conditions'. Even if one is able to identify the correspondence between scales, the second challenge is to establish that those points, related by the mapping ${\cal R}$ are indeed close to each other.

Let us now argue that such strategy indeed exists, and uses in a crucial way the notion of physical, Dirac observables.
The idea is simple: we can use observables to compare physical states at two scales and decide when they are close to each other, namely when the value of such observables is nearby. For that,
we want to construct the family of complete Dirac observables, at every scale
$\l$, and analyze its behavior as $\l\to 0$. 
We say that the effective theories converge as $\l\to 0$ if there is a family of complete Dirac
observables that converge in this limit. Convergence of observables will then induce a flow ${\cal R}$
on the space of theories and we shall then be in position of giving meaning to the limit $\l\to 0$.
 
In the next part we shall review the construction of complete observables from partial observables that 
will be useful for our purposes. 

\subsection{Observables}

Since we are dealing with constrained systems we need to find gauge invariant phase space functions,
i.e. Dirac observables. We shall follow the ideas of Rovelli and Dittrich, who associate to every
pair of phase space functions a one-parameter family of Dirac observables \cite{rovelli,dittrich}. 
We first construct a phase space function $R(q_i,p_i)$ which is a monotonous function of $t$, the evolution
parameter associated to ${\cal C}$. We can choose $R$ as
a function that measures the ``time'' along the gauge orbits, generated by the constraint.
We shall consider another phase space function $F$ and calculate its value when $R=t_0$, the result
$F|_{t_0}$ will then give a Dirac observable. The procedure goes as follows. First, we consider the flow $\alpha^t_C$ 
generated by the constraint ${\cal C}$. For an arbitrary smooth phase space function $F$ it can be calculated as

\be
\alpha^t_C(F)=\sum_{n=0}^{\infty}\frac{t^n}{n!}\{ {\cal C},F\}_n\, ,
\ee
where $\{ {\cal C},F\}_0=F$ and $\{ {\cal C},F\}_{n+1}=\{ {\cal C},\{ {\cal C},F\}_n\}$. Then, one can define
the function
\be
F\vert_{t_0}\equiv \alpha^t_C(F)\vert_{\alpha^t_C(R )=t_0}\, .
\ee
Since $\alpha^t_C(R )=t_0$ is an invertible function, $F\vert_{t_0}$ is a family
of Dirac observables, i.e. $\{ F\vert_{t_0},{\cal C}\} =0$ (For details, see \cite{dittrich}).

In the following sections we shall show that the question of
convergence is a subtle one that one has to consider carefully, in every particular case.  
In our case, the ``time function" $R$ will in general depend on the parameter $\lambda$. As we shall see, one needs to carefully select such time function in order to built meaningful complete observables.
We shall also consider gauge trajectories and their convergence, as well as the relation between the
corresponding reduced phase spaces.

\section{Parametrized Harmonic Oscillator}
\label{sec:3}

In order to illustrate some aspects of convergence of the effective theories let us analyze 
as an example the case of parametrized simple harmonic oscillator (PSHO), which is a reformulation of the
SHO as constrained theory. 

The kinematical phase space $\Gamma_{\rm psho}$ of a 1-dimensional PSHO is four dimensional, and convenient coordinates are $(q,p;\t ,\Pi_\t)$, where $\t$ is the original Newtonian time that has been elevated to a phase space coordinate. The theory is defined by the constraint 
\be 
{\cal C}=\frac{p^2}{2m}+\frac{k}{2}q^2+\Pi_\t\approx 0 \label{sho}\, . 
\ee 
The constrained surface $\bar{\Gamma}\subset \Gamma_{\rm psho}$ is given by solutions to the previous equation.
The `dynamical' solutions for $q$ and $p$ of the corresponding equations of motion 
are periodic functions of $\t$, with constant period $T=2\pi\sqrt{\frac{m}{k}}$. One should, however, have in mind that this `evolution' is a gauge transformation, and the corresponding solutions are the gauge orbits on the constrained surface.

Let us now consider, in the same kinematical phase space $\Gamma_{\rm psho}$, 
the one parameter family of effective theories defined 
by constraints 
\be 
{\cal C}_\l=\frac{k}{2}q^2+\frac{1}{\l^2m}(1-\cos{(\l p)})+\Pi_\t\approx 0 \label{pendulum}\, , 
\ee 
where $p$ is now effectively compactified, since all the dependence on such coordinate is periodic with period
$\frac{2\pi}{\l}$. For convenience, let us assume $p$ is periodic and lies in the interval $p\in [-\frac{\pi}{\l},\frac{\pi}{\l})$. 
In the limit $\l\to 0$ this constraint reduce to (\ref{sho}), and the range of $p$ is the real line. 
The constraint (\ref{pendulum}), for every $\l$ describes a 
`pendulum' (where the periodic variable is $p$ and not $q$), with an energy $-\Pi_\t$, and with the 
formal correspondence given by $p=\l\theta$, $l=\frac{1} 
{\l\sqrt{km}}$ and $g=\frac{\sqrt{km}}{\l m^2}$ 
($\theta$ is an angle, $l$ the length and $g$ is a constant acceleration). 

The gauge orbits, when deparametrized and interpreted as dynamical evolution, exhibit two types of 
trajectories in the phase space. Since $\Pi_\t$ is a Dirac observable, and therefore constant along the gauge orbits, 
one can use it to separate the two class of trajectories. First, there are orbits that correspond to oscillations, 
for $|\Pi_\t|<E_c$, with $E_c$ a {\it critical energy}. The second class of orbits are
rotations for $|\Pi_\t|>E_c$. The quantity $E_c=\frac{2}{\l^2m}$, the critical energy, is a constant on each space
$\bar{\Gamma}_\l$, and diverges in the limit $\l\to 0$. The surface $|\Pi_\t|=E_c$ is where the
corresponding phase space separatrices lie. 
The motion of the pendulum is periodic. For instance $p (\t )$ has a period that depends on $\l$ and $\Pi_\t$
given by 
\be 
T(\l, \Pi_\t)=4\sqrt{\frac{m}{k}}\;F\left(\frac{\pi}{2},\a \right)\, , 
\ee 
where $F$ is the elliptic integral of the first class, and $\a =\sin{\frac{\l p_0}{2}}=\sqrt{-\frac{\Pi_\t} 
{E_c}}$. 
%with $\cos{\l p_0}=1+\l^2mp_\t$. 
For small $\a$ 
$$ 
T(\l, \Pi_\t)=2\pi\sqrt{\frac{m}{k}}\;\left(1+\frac{1}{4}\a^2+\frac{9}{4\cdot 16}\a^4+\ldots \right) 
$$ 
Note that small $\a$ is precisely the regime where the effective theory approximates very well the SHO. We can attain
small $\a$ by fixing the energy $\Pi_\t$ and then making $\l$ sufficiently small. Alternatively, for a fixed $\l$, the approximation is good if we restrict ourselves to energies much lower than $E_c$. 
What is then the procedure to identify orbits for different values of $\lambda$? In this case the answer is simple.
One can, to begin with, fix the observable $\Pi_\t$, and synchronize the initial condition  for $p$ at $\t=\t_0$. With
this prescription, in the plane $(\t,p)$ one can superimpose two orbits, with the same `energy' $\Pi_\t$, with different values of $\l$, that start from the same point $p$ at a given time $\t=\t_0$.

Let us now consider observables in order to pose our convergence conditions.
For Dirac observables in both theories we can choose $\Pi_\t$, as it is already a Dirac observable and therefore, a constant of the motion. The other natural choice is to consider   
the family of complete observables, $p|_{\t_0}$, corresponding to `the value of $p$ for fixed $\t =\t_0$'. 
We can therefore choose 
$p$ and $\t$ as partial observables and calculate their flow. 

In the case of the classical SHO we easily  obtain 
$$ 
\alpha^t_C(\t )=\t -t\, ,\ \ \ \ \alpha^t_C(p)=p\cos{\omega t}+\frac{k}{\omega}q\sin{\omega t}\, , 
$$ 
then the observable becomes
\be 
p|_{\t_0}=p\cos{(\omega (\t -\t_0))}+\frac{k}{\omega}q\sin{(\omega (\t -\t_0))}\, . 
\ee 
In the case of the effective theories, the flow $\alpha^t_{C_\l}(p)$ can not be obtained in closed 
form, but nevertheless we can still draw some conclusions about the behavior of the corresponding 
family of Dirac observables, for different values of $\l$. Let us first remark that the 
difference between the values of $p|_{\t_0}$ and $p_\l|_{\t_0}$ (the notation $p_\l$ indicates 
that we are in the effective theory labeled by $\l$), for $|\Pi_\t|<E_c$ fixed, is always bounded 
\be 
\Bigl| p|_{\t_0} -p_\l|_{\t_0}\Bigr|\le \sqrt{-2m\Pi_\t}+\frac{1}{\l}\arccos{(1+\l^2 m\Pi_\t )}\, , 
\ee 
Note that this inequality provides a bound, but it does {\it not} tend to 0 as $\l\to 0$. Nevertheless, 
the left hand side tends to zero pointwise when  $\l\to 0$, due to the fact that dynamical trajectories approach each other in this limit. However, this convergence is not uniform. If we compare $p_{\l_1}|_{\t}$ 
with $p_{\l_2}|_{\t}$ we notice that their difference is not periodic in $\t$, 
so if we start with two nearby values of $p$, for $\t$ and $\Pi_\t$ fixed, on two different 
scales, they do not stay nearby as $\t$ changes, due to the fact that the period of $p_\l (\t)$, 
depends on $\l$. Thus, the naive strategy of constructing observables, starting from the same function
$\t$ in all theories labeled by $\l$, does not yield observables that are close to each other.
Let us now define slightly different observables, where the idea is to try to {\it synchronize} the oscillation
for all values of $\l$.

One can 
synchronize measurements in different effective theories (with different $\l$) as well as the SHO with an 
effective theory for 
some given $\l$, only if the solutions in all the theories have the same period. In order to achieve 
that we introduce a new parameter, in all the effective theories, defined by
\be 
\t_\l =2\pi\sqrt{\frac{m}{k}}\;\frac{\t}{T(\l, \Pi_\t)}   
\ee 
Notice that $\t_\l\to\t$ as $\l\to 0$. Now, as a complete set of Dirac observables in effective theories 
we can consider $\Pi_\t$, and
$p|_{\t_\l}$, the value of $p$ for $\t_\l$ fixed. We can now test the convergence of these observables
by noting that
%It is easy to show that 
\be 
\Bigl| p|_{\t =\t_0} -p|_{\t_\l =\t_0}\Bigr|\le \Bigl|\sqrt{-2m\Pi_\t}-\frac{1}{\l}\arccos{(1+\l^2 m\Pi_\t )}\Bigr|\, , 
\ee 
which is uniformly bounded and tends to 0 as $\l\to 0$. It is important to note that in the previous equation we are evaluating the flux of the observable $p$, at the point where the fluxes of the partial observables $\t$ and $\t_\l$, along the orbit generated by ${\cal C}$, take the value $\t_0$.  Also, we have  used the shorthand notation $\t_\l =\t_0$ to denote $\a_{C}^t(\tau_\l)=\t_0$.
We shall use this notation in what follows. As we have shown,
this new set of Dirac observables converge in a uniform way as $\l\to 0$ and, therefore, provide a precise
sense in which the effective theories converge to the classical one, as $\l\to 0$. 

Let us end this section by making some remarks about the reduced phase space of the PSHO and the effective 
theories. 
The kinematical phase space in all of these theories is given by $\Gamma = \mathbb{R}^4$, 
with the coordinates $(q,p,\t ,\Pi_\t)$. The constraints define a family of hypersurface $\bar{\Gamma}_\l\subset \Gamma$, and the resulting induced pre-symplectic form $\bar{\Omega}$ ( 
$\bar{\Omega}_\l$) 
has degenerate directions corresponding to the Hamiltonian vector fields of the constraint. 
Every gauge orbit is represented by a point in the 
reduced phase space $\hat{\Gamma}_\l$. An alternative description is to consider a {\it gauge fixing} condition, in which one selects one point of each equivalence class. For that purpose we select the following gauge fixing condition: 
$\t =\t_0$, with $\t_0$ a constant. It is easy to show that this is an acceptable global gauge choice. 
The reduced phase space can then be parametrized in various ways. For instance, 
we can choose $(q,p)$, since the corresponding (non-degenerate) symplectic structure is independent of $\l$, 
$$\hat{\Omega}_\l=\d p\wedge \d q\, .$$

It might appear that we have arrived to a common description for all $\hat{\Gamma}_\l$, given that neither the coordinates nor the symplectic structure depend on $\l$. However, one has to be careful with the periodicity
conditions introduced by $\l$.
In the PSHO theory, Dirac observables are arbitrary functions 
in the reduced phase space, while in the effective theories they have to be periodic functions of $p$
with a period $\frac{2\pi}{\l}$. Thus, even in this picture, we are forced to consider $\l$-dependent observables
(functions on $\hat{\Gamma}_\l$) to talk about convergence, as we have done in previous parts of this section.

%%%%%%%%%%%%%%%%%%%%%%%%%%%%%%%%%%%%%%%%%%%%%%%%%%%%%%%%%%%%%%%%%%%%%%

\section{$k$=0 Isotropic Loop Cosmology}
\label{sec:4}

Our next example will be a simple isotropic and homogeneous cosmological model that has received
some of attention given that it can be solved both classical and quantum mechanically, for
arbitrary values of $\l$ in loop quantum cosmology. This will allow us to make our 
consideration precise and analyze in detail the issue of convergence. The model in question 
is a $k=0$ FRW cosmological model coupled to a massless scalar field. For details of the 
model see \cite{slqc}. 
 %In order to have
%well defined quantities all integrals should be restricted to a fiducial cell ${\cal V}$.
%We can introduce a flat fiducial metric $\mathring{q}_{ij}$ and $\mathring{V}$ is
%the volume of ${\cal V}$ with respect to this metric. The physical metric is given by a
%lapse function $N(\t )$ and the scale factor $a(\t )$. We choose $N(\t )=1$, so the 'natural'
%evolution parameter is proper time.
The classical phase space for the gravitational sector is described by two
variables $\b$ and $V$, where, on shell, $\b$ is proportional to the Hubble parameter,
$\b =\gamma\frac{\dot{a}}{a}$ ($\gamma$ is the Barbero-Immirzi parameter, and $a$ is the scale factor) and $V$ 
is the physical volume of the cell ${\cal V}$. %$V=\mathring{V}a^3$. 
Note that $\b\in (-\infty ,\infty )$
and $V\ge 0$. The pair conjugate variables satisfy $\{ \b ,V\} =4\pi G\gamma\, .$ The matter sector is given by
a scalar field $\phi$ and it momentum $p_\phi$, such that 
$\{\phi ,p_{\phi}\}=1$. 

\subsection{Classical theory}

The classical system is subject to the Hamiltonian constraint:
\be
{\cal C}=-\frac{3}{8\pi G\gamma^2}V\b^2+\frac{{p_{\phi}}^2}{2V}% +U(\phi )V
\, .
\ee
%with $U(\phi)$ is a potential.
%The curve $\b =0$ (for $p_{\phi}\ne 0$ and $U(\phi )\ge 0$) does not belong to ${\cal C}$.
The `equations of motion', corresponding to the gauge motions generated by the
constraint ${\cal C}$ are
$$\dot\b 
%\approx \frac{3}{\l^2} \sin^2{(\l\b)}-\frac{2\kappa^2}{3}U(\phi )
\approx -4\pi G\gamma\frac{p_{\phi}^2}{V^2}\, , \quad \dot V=\frac{3}{\g}V\b\, , 
%$$
%$$
\quad \dot\phi =\frac{p_{\phi}}{V}\, ,\quad \dot{p}_\phi = 0\, .%-V\frac{dU(\phi )}{d\phi}\, .
$$
%The energy density and the pressure of the scalar field are given by
%\be
%\rho =\frac{p_{\phi}^2}{2V^2}+U(\phi )\, , \ \ \ \ p=\frac{p_{\phi}^2}{2V^2}-U(\phi )\, ,
%\ee
from which one can obtain the Friedman equation that contains information about ``dynamics"
\be
\left(\frac{\dot{a}}{a}\right)^2=\left(\frac{\dot{V}}{3V}\right)^2 = \frac{8\pi G}{3}\rho\, ,
\ee
%so the only classical turning point (where $\dot{V}=0$) occurs when $\rho =0$ and for a
%standard scalar field this can happen only for negative potential, leading
%to a recollapse.
where $\rho$ is the scalar field density.
Here the derivative $\dot{}:=\frac{\d\,}{\d\tau}$, is taken with respect to the cosmological proper time $\t$,
that happens to be the affine parameter of the gauge orbits generated by ${\cal C}$ (or, alternatively, for
lapse $N=1$). In order to gain some insight into the structure of the gauge orbits, let us study some aspects of
the solutions of the dynamics equations. 
Note that,  $\phi (\t )$ and $\b (\t )$ are monotonic functions,
so both can be used as an internal relational time. The
momentum $p_{\phi}$ is a Dirac observable and therefore, a constant of motion (we choose $p_{\phi}>0$).
In the classical theory the solutions of the Hamilton's equations
are not defined for $\tau =0$, so there are two branches, corresponding to the sign of $\beta$, 
$$
V(\t )=({\rm sgn}\b )\,\k p_{\phi}\,\t \, , \qquad \b(\t )=\frac{\gamma}{3\t}\, ,$$
$$ 
\qquad \phi (\t )=\phi_0+\frac{({\rm sgn}\b )}{\k}\ln{\frac{\t}{\t_0}}\, ,
$$
where $\k =\sqrt{12\pi G}$,
an integration constant is fixed with the condition $V(\t )\to 0$ as $\t\to 0$, and $\phi_0 =\phi (\t_0)$. 
Note that the constrained surface $\bar{\Gamma}$ is disconnected in four connected components (labeled by the signs
of $\b$ and $p_\phi$). Choosing $p_\phi>0$ does not remove any generality (we can recover those
branches in a trivial manner), so we are then left with two connected components, labeled by sgn$\b$.
Gauge orbits with the space-time interpretation of an expanding universe correspond to ${\rm sgn}\b >0$, while the contracting branch corresponds to ${\rm sgn}\b <0$. 

Our objective is to find suitable observables that will allow us to study the issue of convergence. Therefore,
we need to understand the dynamics generated by the constraint in terms of an internal dynamics, where the 
external, unphysical parameter (in this case, $\tau$), is not present. Let us then explore the different possibilities we have at our disposal. 
This discussion will then be taken over to the effective description.
Since, $\dot{\b}\le 0$ one can choose $\b$ as a relational time in the classical
theory, and consider the evolution with respect to $\b$. The advantage of this
election is that no external time variable is needed. Every trajectory,
that corresponds to $\b >0 $, expands forever, where the big bang corresponds to
$\b=\infty$ and the volume increases without bound as $\b\to 0$. 

The scalar field can also be used as an internal time variable, the volume as a function of
$\phi$ is given by
\be
V(\phi )= V_0 e^{\k ({\rm sgn}\b ) (\phi -\phi_0)}\, ,
\ee
where $V_0=V(\phi_0)$.
These two possibilities will be further explored when constructing complete observables. Let us now considered
the $\l$-dependent descriptions.

\subsection{Effective theories}

In the solvable LQC model, the effective dynamics with non-zero parameter $\l$ can be seen as containing corrections
% is obtained as
%an approximation of the expected value of the quantum constraint
%over semi-classical states and leads to the corrections of the standard cosmology
due to quantum geometry effects.
The parameter $\lambda$ 
is associated to a fundamental granularity of quantum geometry.
It is also possible to consider $\lambda$ as a regulator in the same   
spirit as those used in quantum field theory, it   
appears in the model as the prescription of a regular lattice in the   
real line. The effective theory at scale $\l$ is determined by the constraint that takes the form \cite{slqc}
\be
{\cal C}_{\l}=-\frac{3}{8\pi G\g^2}\frac{1}{\l^2}V\sin^2{(\l\b)}+\frac{{p_{\phi}}^2}{2V}%+U(\phi )V
\, ,
\ee
\noindent where $\b$ can be seen as being compactified, taking values in $\b\in [-\frac{\pi}{2\l},\frac{\pi}{2\l})$.

The only equation of motion different from the classical one (on the constraint surface) is
\be
\dot V=\frac{3}{\g\l}V\sin{(\l\b)}\cos{(\l\b)}\, ,
\ee
leading to the modified Friedman equation 
\be
\left(\frac{\dot{a}}{a}\right)^2=\left(\frac{\dot{V}}{3V}\right)^2=\frac{8\pi G}{3}\,\rho\, \left(1-\frac{\rho}{\rho_b}\right)\, ,
\ee
where $\rho_b=\frac{9}{2\kappa^2}\frac{1}{\l^2}$ is the scalar field density at the bounce.
At each scale there are quantum turning points at $\b =\pm\frac{\pi}{2\l}$, where $\dot{V}=0$,
and it corresponds to a bounce. Note that, at the bounce
$$\ddot V\vert_{\beta =\frac{\pi}{2\l}}=2\kappa^2V\rho_b >0\, ,$$
so the bounce corresponds to a minimum of volume.

In the case of effective theories the proper time appears as a natural choice for an
evolution parameter. In the corresponding quantum theories there is no notion of a proper
time, nevertheless one can choose a relational time variable, as a function on the
reduced phase space, with respect to which the quantum dynamics can be described.

Since, $\dot{\b}\le 0$ one can choose $\b$ as a relational time in the effective
theories, and consider the evolution with respect to $\b$. The advantage of this
election is that no external time variable is needed. Every trajectory,
that corresponds to $\b >0$, has a bounce
at $\b =\frac{\pi}{2\l}$, and this value tends to infinity as $\lambda\to 0$. If
we are interested in the analysis of the continuum limit of the effective theories
we should pay a special attention to the behavior of the solutions at the bounce.
To be able to investigate it we need an evolution parameter that does not diverge
at the bounce as $\lambda\to 0$. So, one possibility is to define
$$
\tau_{\l} (\b)=\frac{\g\l^2}{3}\int^{\frac{\pi}{2\l}}_{\b}{\frac{\d u}{\sin^2{\l u}}}
$$
as a function on the constrained surface $\bar{\Gamma}_\l$, such that $\tau (\frac{\pi}{2\l})=0$.
The analogous construction should be performed for
$\b <0$, as well. The result is exactly the proper time. The conceptual difference is that
$\tau_\l (\b)$ is to be seen as a function on $\bar{\Gamma}_\l$, that has
been constructed in search of a consistent description from the Hamiltonian perspective. 
Interestingly, the end result coincides
with the original parameter $\t$ that was the affine parameter of the original constraint (with $N=1$),
and has the space-time interpretation of cosmic proper time. 

%\subsection{Massless scalar field}
In the effective theories, we consider the interval $\b\in [-\frac{\pi}{2\l},\frac{\pi}{2\l})$.\footnote{Recall that all the functions and observables in $\bar{\Gamma}_\l$ are periodic in $\beta$ with period $\pi/\l$. It is then totally equivalent to regard the coordinate as compactified on a circle.}
The solutions are defined for every $\t$ and are given by
\be
\cot{\l\b}=\frac{3\t}{\g\l}\, ,\ \ \
V_{\l}(\t )=\frac{\k}{3}p_{\phi}\sqrt{\gamma^2\l^2+9\t^2}\, ,
\ee
and
\be
\phi_{\l} (\t )=\phi_0 +\l\varphi +\frac{1}{\k}\ln{\frac{3\t+\sqrt{\gamma^2\l^2+9\t^2}}
{3\t_0+\sqrt{\gamma^2\l^2+9\t_0^2}}}\, ,
\ee
so that $\phi_{\l}(\t_0)=\phi_0 +\l\varphi$
and the initial condition approaches the classical one (for $\t=\t_0$) as $\l\to 0$.
Note that $\phi_{\l}(0)\to\frac{{\rm sgn}\b}{\kappa}\ln{\l}$ as $\l\to 0$.

In order to gain more intuition about the behavior of the effective solutions and their relation to their 
classical counterpart, let us consider the following two observations. First,
we can consider the proper time needed to expand from $V(0)$ to some fixed $\tilde{V}$. 
If we denote by $\tau_0$ the proper time it would take in the classical
theory, then in the effective theory given by $\l$ the corresponding proper time is
given by $\tau_{\l}=\sqrt{\tau_0^2-\frac{\gamma^2\l^2}{9}}$. Note that for late times
after the bounce, namely when $\t_0\gg \gamma\l$, the age of the universe in the classical and 
effective theories are practically the same. Second, we can ask the following question.
Given a matter density, say, at late times after the bounce, can we find the age of the universe?
That is, what is the proper time elapsed from the bounce to that instant. The answer is strikingly 
simple.
In the effective theory there is a maximal density given by $\rho_b=\frac{p_{\phi}^2}{2V(0)^2}$. The proper time needed to reach $\rho =\frac{\rho_b}{D}$ (with $D >1$) is then
$\tau =\frac{1}{3}\gamma\l\sqrt{D -1}$. Finally, note that, for any given 
value of $p_\phi$, $V_{\l}(\t )$, $\b_{\l} (\t )$ and $\phi_{\l} (\t )$
converge uniformly in the limit $\l\to 0$ to an expanding branch of the classical trajectory 
for $\t> 0$, and to a contracting branch for $\t <0$. 

For all effective theories the bounce occurs at $\t =0$, at this point
$$
V_{\l}(0)=\frac{\k}{3}p_{\phi}\gamma\l\, ,\qquad \dot V_\l (0)=0\, ,\qquad 
\ddot V_\l (0)=\frac{3\k p_\phi }{\gamma\l} \, .
$$
Note that, in the limit $\l\to 0$, $V_{\l}(0)\to 0$ and $\ddot V_\l (0)
\to\infty$. That is, the hyperbolas in the plane $(\t,V)$, become degenerate in the limit $\l\to 0$ to 
a curve proportional to $|\t|$, that has a spike at the origin.

Let us now consider, as we did for the classical theory, an intrinsic description of the dynamics.
In order to solve $V$ as a function of $\phi$, note that
$$
\frac{dV}{d\phi}=\kappa ({\rm sgn}\b )\sqrt{V^2-\alpha^2}\, ,$$
where $\alpha =\frac{1}{3}\gamma\l\kappa p_\phi$. The solution is of the form
\be
V_{\l}(\phi )=V_+e^{\k ({\rm sgn}\b )  (\phi -\phi (\t_0))}
+V_-e^{-\k ({\rm sgn}\b )(\phi -\phi (\t_0))}\, ,
\ee
where $V_+=\frac{1}{2}(V_0+\sqrt{V_0^2-\alpha^2})$ and $V_-=\frac{\alpha^2}{4}(V_+)^{-1}$,
where $V_0=V(\phi (\t_0))$.
Note that $V_+\to V_0$ and $V_-\to 0$ as $\l\to 0$. 
The convergence of $V_\l (\phi )$ as $\l\to 0$ is not uniform, because $\phi_{\l}$ at 
the bounce tends to $\mp\infty$ in this limit. Thus, describing the intrinsic dynamics in terms
of the scalar field $\phi$ is not convenient for taking the limit. In other words, there is
no continuum limit with respect to the internal time $\phi$. Note that this result is consistent with previous claims \cite{slqc,CVZ3}.

%We illustrate these convergence results at the following figures.
\begin{figure}
 \centering
  \epsfig{file=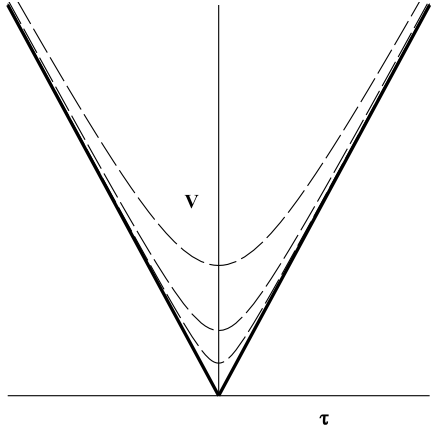,scale=0.3}
\caption{Volume as a function of proper time for different values of $\lambda$ and 
fixed $p_\phi$. The solid lines
corresponds to the classical expanding trajectory, for $\t >0$ and the contracting one for $\t <0$.}
\end{figure}

\begin{figure}
 \centering
  \epsfig{file=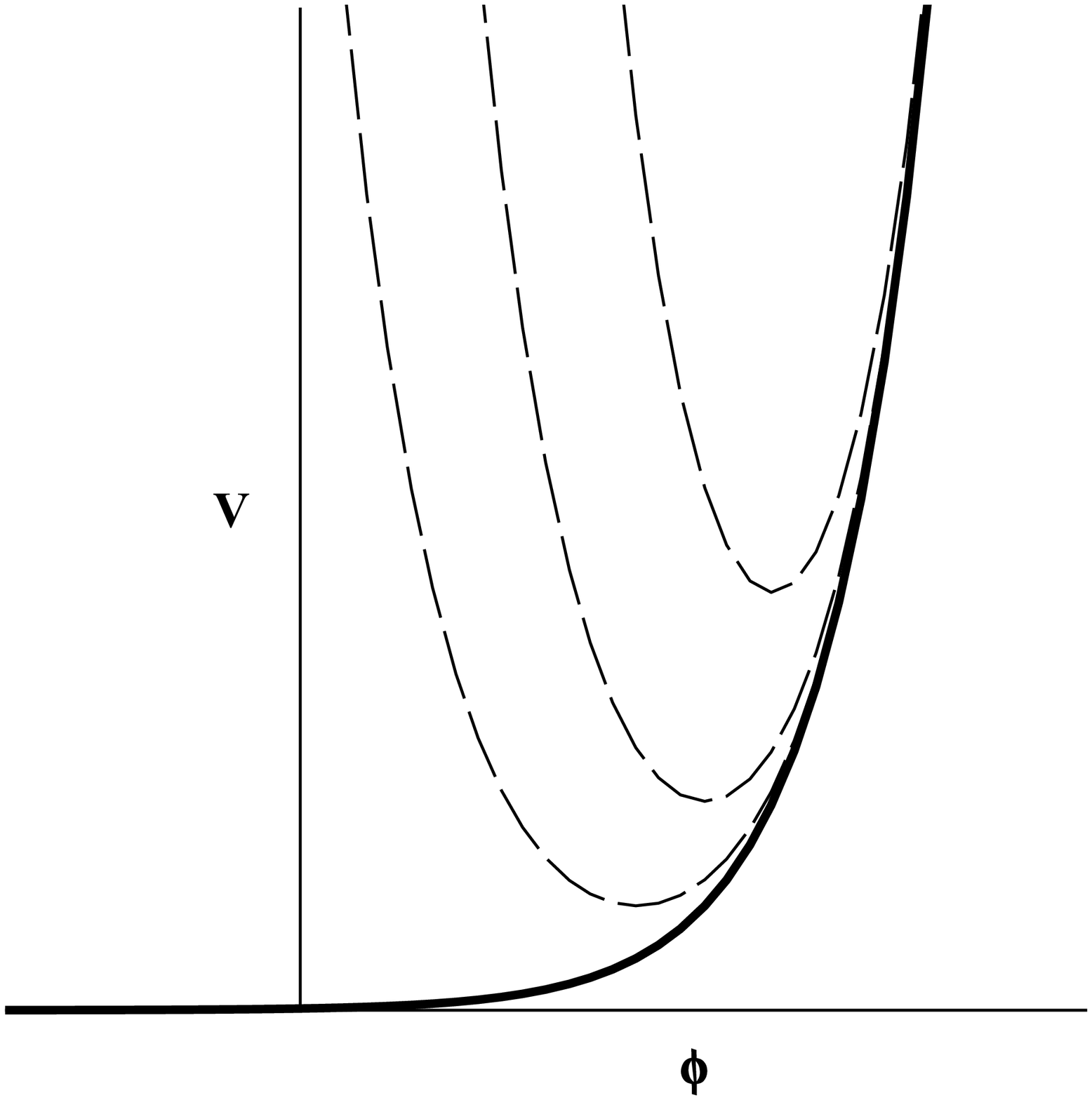,scale=0.3}
\caption{Volume as a function of scalar field for different values of $\lambda$ and fixed $p_\phi$, for $\t_0>0$. 
The solid curve corresponds to the classical expanding trajectory.}
\end{figure}

As we mentioned $\b$ can also be used as a relational time, and $V=V(\b )$ is determined from
${\cal C}_{\l}\approx 0$.

\begin{figure}
\centering
\epsfig{file=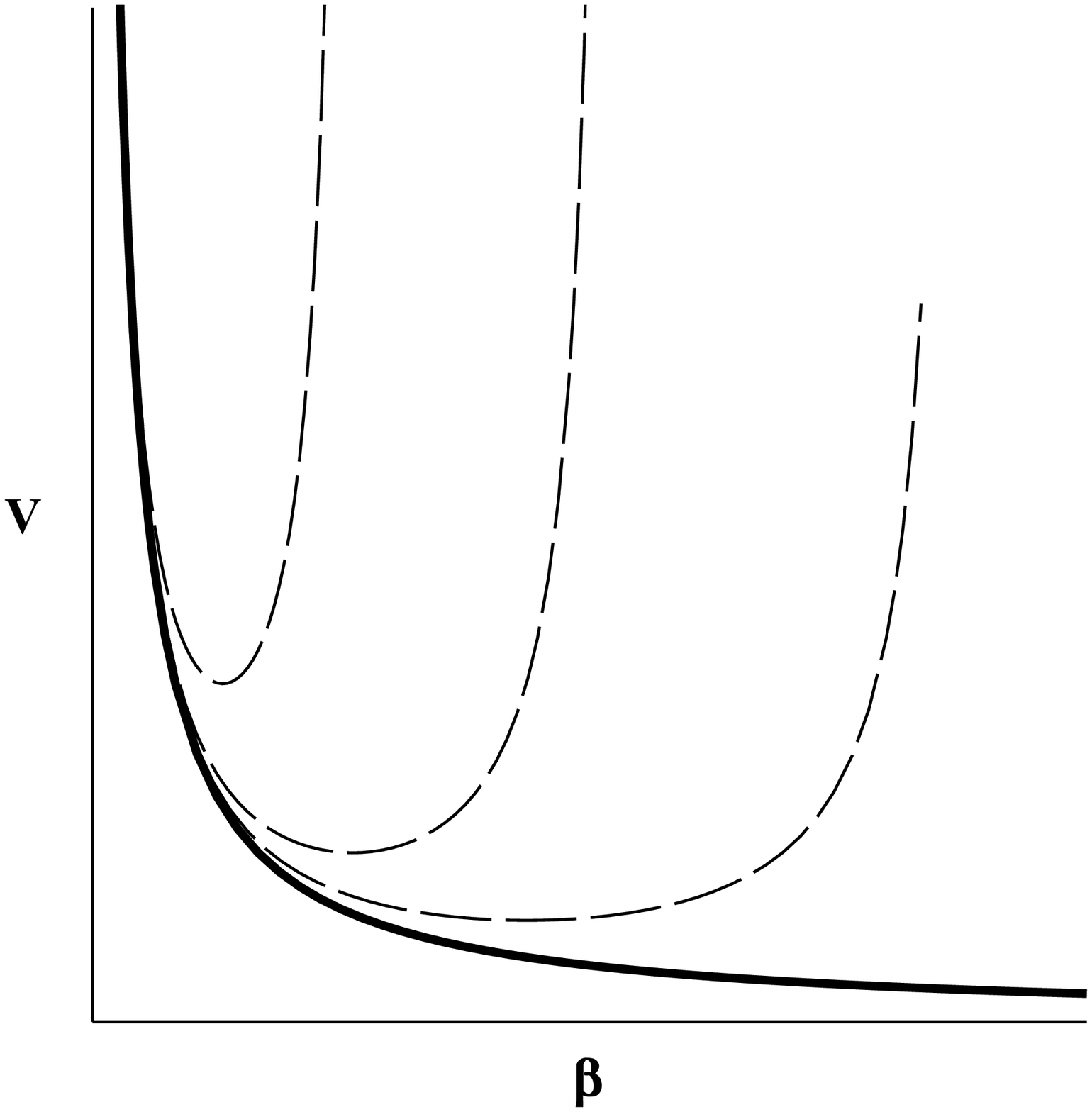,scale=0.3}
\caption{Volume as a function of $\b >0$ for different values of $\lambda$ and fixed $p_\phi$. The solid curve
corresponds to the classical expanding trajectory.}
\end{figure}

\begin{figure}
 \centering
  \epsfig{file=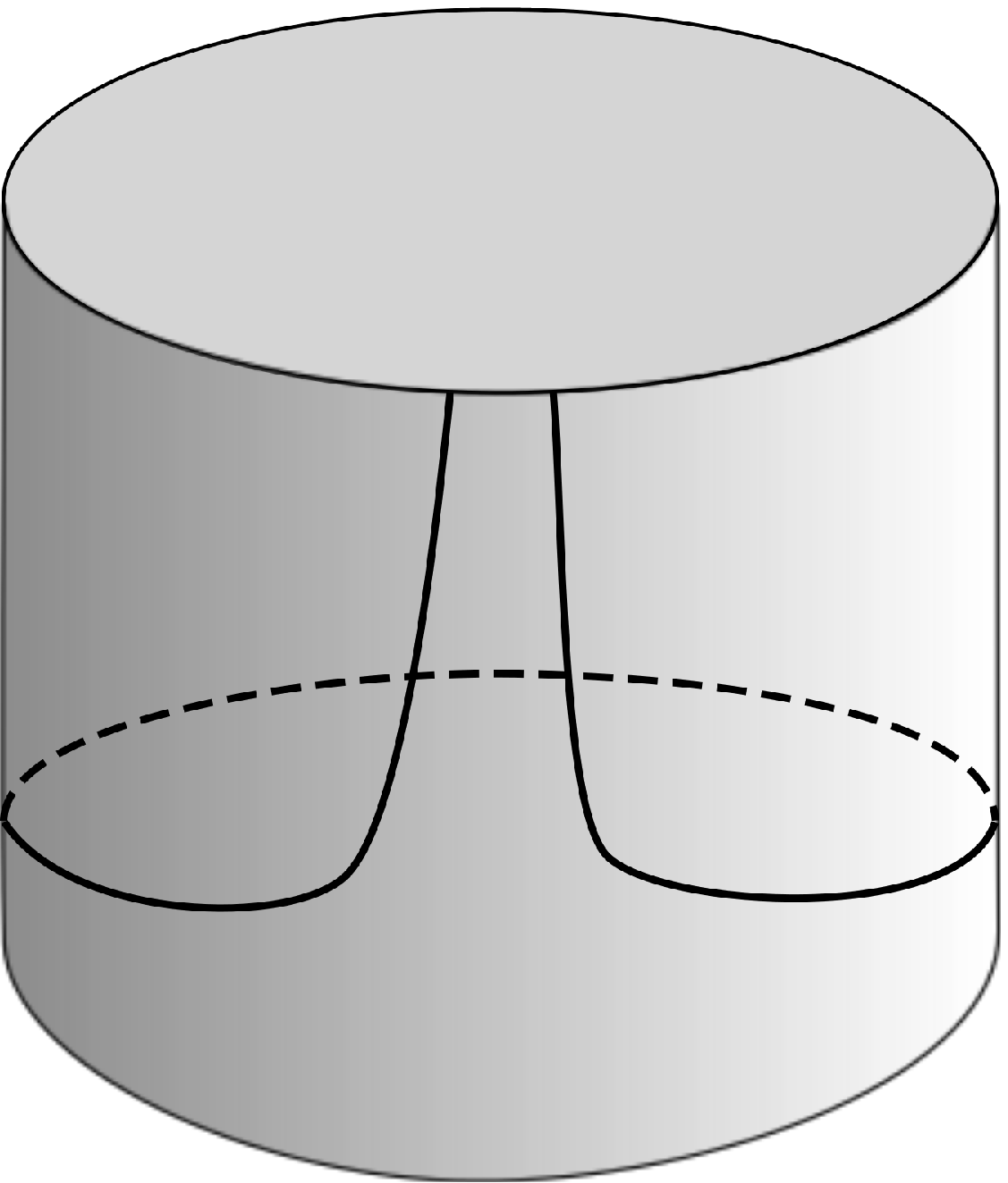,scale=0.3}
\caption{Volume as a function of compactified $\b$ for fixed $\lambda$ and $p_\phi$. }
\end{figure}

The convergence results are easy to see, for any time chosen to study the evolution of the system. However, the 
convergence is in general not uniform. Since the high density regime (near $\rho_b$) is of special interest, one 
should consider the evolution parameter that yields the better convergence behavior at the bounce. In this case the 
preferred choice is the proper time, because the bounce occurs for $\t_\l =0$, for every effective theory and, in the 
limit $\l\to 0$, the volume $V_\l (0)$ is well defined, allowing us to connect the two branches of the classical 
theory. The theory obtained in this limit is well defined before and after the bounce.

All this consideration have given us an intuitive understanding of the behavior of solutions and
to the nature of the convergence. Let us now formalize these results by considering complete Dirac
observables in the spirit of the rest of the sections.

\subsection{Partial and complete observables}

In this part we shall construct complete Dirac observables for both the classical theory and the family of
effective theories.

Let us start by considering 
$V$ and $\phi$ as two partial observables of the theory and construct a family of complete (Dirac)
observables $V\vert_{\phi_0}$ labeled by a parameter $\phi_0$. In order to calculate the flow of $\phi$
we have to calculate $\{\phi ,{\cal C}\}_n$. The expressions turn out to be complicated because of
the presence of the $V^{-1}$ term in ${\cal C}$. We can simplify these expressions if we first redefine 
the classical constraint as $\tilde{C}:=V{\cal C}$, and calculate the flow of partial observables
with respect to the new constraint $\tilde{C}$. This corresponds to a change in the lapse function, with 
the new choice being $N(t)=a^3$ \cite{slqc}.
Then, 
\be
\alpha^t_{\tilde{C}}(\phi )=\phi -p_{\phi}t\, , \qquad \alpha^t_{\tilde C}(V)=Ve^{-\frac{3}{\gamma}V\b t}\, .
\ee
We see that $\phi$ is a good parametrization of the gauge orbits. The family of complete 
observables is then given by
\be
V\vert_{\phi_0}=Ve^{-\kappa ({\rm sgn}\b)(\phi -\phi_0)}\, .
\ee
On shell, one obtains $V\vert_{\phi_0}=V_0$.

If instead of considering the plane $(\phi ,V)$ we choose $(\tau , V)$ as the partial observables with
\be
\t = ({\rm sgn}\b )e^{\kappa ({\rm sgn}\b )\phi}\, ,
\ee
 defined as a proper time (on $\bar{\Gamma}$), we find that
\be
V\vert_{\t =\t_0}=V\frac{\t_0}{\t}\, .
\ee

In the case of the effective theories we proceed as above and redefine the constraint as $\tilde{C}_\l =VC_\l$ and calculate the flows of the partial observables. The flow of $\phi$ is the same as above, while
\be
\alpha^t_{\tilde{C}_\l}(V)=V\left[ \sinh{(-\frac{3}{\gamma\l}Vt\sin{\l\b})}\cos{\l\b} + 
\cosh{(\frac{3}{\gamma\l}Vt\sin{\l\b})}\right]\, ,
\ee
resulting in the following family of Dirac observables
\be
V_\l\vert_{\phi_0}=V\bigl[\sinh{(-\kappa ({\rm sgn}\b)(\phi -\phi_0) })\,\cos{\l\b} + 
\cosh{(\kappa ({\rm sgn}\b) (\phi -\phi_0))}\bigr]\, .
\ee

If we now define a proper time as 
\be
\t_\l =\frac{\gamma\l}{3}\sinh{(\kappa ({\rm sgn}\b)\phi +\ln{\frac{3}{\gamma\l}})}\, ,
\ee
we find that
\be
V_\l\vert_{\t_\l =\t_0}=\frac{1}{2}V\bigl[ (z-\frac{1}{z})\,\cos{\l\b}+(z+\frac{1}{z})\bigr]\, ,
\ee
with $z=(3\t_0+\sqrt{\gamma^2\l^2+9\t_0^2})(3\t_\l +\sqrt{\gamma^2\l^2+9\t_\l^2})^{-1}$. 

We can now compare Dirac observables in the classical and the effective theory, for the different choices
of internal dynamics we have considered. In the first case we compare the volumes
for fixed value of a scalar field
\be
V_\l\vert_{\phi_0}-V\vert_{\phi_0}=\frac{1}{2}(\sqrt{V_0^ 2-\a^2}-V_0)e^{-\k (\phi_0-\phi )({\rm sgn}\b )}
+\frac{1}{2}\frac{\a^2}{V_0+\sqrt{V_0^ 2-\a^2}}e^{\k (\phi_0-\phi )({\rm sgn}\b )}\, ,
\ee
we see that this expression tends to 0 as $\l\to 0$, but the convergence is non-uniform, as we could have anticipated
given the result of the previous part.

On the other hand, for the observable corresponding to volume at a fixed value of a $\l$-dependent proper time $\t_\l$, we obtain
\be
V_\l\vert_{\t_\l =\t_0}-V\vert_{\t =\t_0}=\frac{\k}{3}p_{\phi}(\sqrt{9\t_0^2+\g^2\l^2}-3|\t_0|)\, ,
\ee
which is uniformly bounded. This proves that there is indeed a `continuum limit'.
Thus, we see that only with the choice of a $\l$-dependent time parameter,
the corresponding complete observables exhibit convergence in the $\l \to 0$ limit. This is exactly the
same result that we found for the case of the parametrized oscillator, where the original choice of
($\l$-independent) Newtonian time did not exhibit convergence. While we can not conclude that this has to be the case,
these two examples provide strong evidence for the claim that one needs to properly select a $\lambda$ dependent
internal time, for the intrinsic dynamics contained in constrained systems to exhibit convergence properties.

Now that we have established the existence of a continuum limit, the obvious question is: what is the limit? Does it 
correspond to the classical theory? The answer is simple: No. As the detailed analysis of the previous part shows, the 
limiting dynamics corresponds to a universe that is recollapsing following the classical dynamics, and such that the 
trajectory can be continued continuously to an expanding one, following again the classical dynamics. One should 
however note that the `limiting dynamics' is non-differentiable at the 
origin, so it does not obey `reasonable' equations of motion (the acceleration $\ddot{V}$ diverges at the origin 
$\t$=0). This non-smooth dynamics of the limiting theory at the origin can informally be called a `singular bounce', 
to distinguish it from the regular bounce exhibited when $\l > 0$.

%\subsection{Reduced phase space}

Let us end this section by making some remarks regarding the reduced phase space point of view of the system.
The kinematical phase space in the FRW theory coupled with a massless
scalar field is $\Gamma = \mathbb{R}^4$, with  coordinates $(V,\b ,\phi ,p_\phi )$. 
The constraints in effective theories define a 1-parameter family of hypersurfaces 
$\bar{\Gamma}_\l$. The resulting induced pre-symplectic form $\bar{\Omega}_\l$
has degenerate directions corresponding to the Hamiltonian vector fields of the constraint.
For any fixed $p_\phi$ there are two gauge orbits in the classical theory, for two different signs of
$\b$. On the other hand in the effective theory $\b$ is compactified and, for  every fixed $p_\phi$, there
corresponds only one gauge orbit.

Every gauge orbit is represented by a point in the
reduced phase space which can be equivalently obtained by gauge fixing. For that purpose we choose $\phi =\Phi$,
with $\Phi$ a constant, that is good global choice of gauge. 
The reduced phase space $\hat{\Gamma}_\l$ can be parametrized in various ways. We choose $(V,\b )$, because the corresponding $\l$-dependent symplectic structure is actually the same as in the classical
theory \cite{CM}
$$\hat{\Omega}_\l=\frac{1}{4\pi G\gamma}\;\d V\wedge \d\b$$
Just as we saw in the case of the parametrized oscillator,
in the classical theory $\b\in (-\infty ,\infty)$ while in the effective theory $\b$ can be defined as
living in a finite interval
$\b\in [-\frac{\pi}{2\l},\frac{\pi}{2\l})$, with the Dirac observables arbitrary functions
in the reduced phase space.
It is however more convenient to extend the domain of $\b$ in the effective
theory to $(-\infty ,\infty )$ and consider only periodic functions with a period $\frac{\pi}{\l}$. In this sense,
one is forced to consider $\l$ dependent observables to study the issue of convergence.

It is now straightforward to see the continuum limit from the perspective of the reduced 
phase space.
In the classical theory every point on $\hat{\Gamma}$ represents an expanding or contracting 
solution. On the other
hand, a point on  $\hat{\Gamma}_\l$ represents a solution that contracts and then expands, 
approaching in both
very early and very late times one of the classical solutions. Thus in the limit, what we 
have is that the series of points $x_\l\in\hat{\Gamma}_\l$ tend not to one point on 
$\hat{\Gamma}$, but to {\it two} points, corresponding to the two solutions that are 
connected by the singular bounce. Thus, there is an essential discontinuity in the limit. 
This is perfectly consistent with our discussion in the main part of this section where we found the continuum limit exists but is discontinuous.

\section{Discussion}
\label{sec:5}

We have considered one parameter families of classical theories labeled by a parameter $\l$ with the interpretation of 
a discreteness parameter. These theories arise as ``effective", semiclassical descriptions of ``polymer quantum 
theories", which are constructed by considering non-standard quantum representations of the canonical commutation 
relations. We have focused our attention on totally constrained systems and have given answers to the following 
question: Can one make sense of a {\it flow} along $\l$ and of the corresponding {\it continuum limit} $\l\to 0$? In 
answering this question we have proceeded in several steps. The first one was to give a precise formulation of what 
exactly taking this limit means, and what criteria we need to impose to talk about convergence of such limit. We put 
forward a detailed prescription in which complete Dirac observables play a fundamental role. Given that at different 
scales $\l$ one is comparing states defined on different spaces, it is only through $\l$-dependent observables that one 
can impose convergence criteria. As we saw, if we are interested in looking at the convergence of `dynamical 
trajectories', one first needs to introduce an internal, relational dynamics for such constrained systems. The frozen 
description provided by the reduced phase space perspective does not provide such scenario. One needs to define an 
internal phase space dependent time function with respect to which the dynamics unravels. In doing so, we learned that 
we could not make any choice that provides a relational dynamics. Only with an appropriate choice of internal time, 
the limit $\l\to 0$ can be defined.

We studied two examples in full detail. The first one was a parametrized harmonic oscillator. In the original, 
`classical system' one has a good understanding of the dynamics contained within the `frozen' dynamics of the 
constrained system. One can de-parametrize the system and recover the original Newtonian time with respect to which 
the system oscillates at a constant period. The family of effective theories are formally equivalent to a pendulum, 
with the oscillator as the $\l\to 0$ limit. What we saw is that one could easily define the $\l$ flow, but the orbits 
were not synchronized when described by a common time function. In order to achieve convergence, one was forced to 
define a new internal time that had a highly non-trivial dependence on $\lambda$, in order to synchronize all the 
pendulae. Only when the dynamics at all scales was synchronized, one could see that there was a uniform convergence to 
the standard harmonic oscillator.

The second example came from loop quantum cosmology. The systems correspond to a $k$=0 FRW cosmology with a mass-less 
scalar field. Classically, all solutions have either a singularity to the past and expand forever, or are contracting 
with a big crunch singularity to the future. On the contrary, all solution for the $\lambda>0$ effective theories 
posses a bounce where the universe transitions from a contracting phase to an expanding one. Both at early and late 
times the effective dynamics approaches one of the classical contracting or expanding solutions. In this case, the 
scalar field $\phi$ provides a natural internal time with respect to which the dynamics can be described. As we saw in 
detail, this choice, however, does not behave well as $\lambda\to 0$. There is another internal coordinate of phase 
space with respect to which one can describe the dynamics for $\lambda\neq 0$, but it is also ill behaved in the 
limit. Interestingly, we saw that one {\it can} define an internal time with respect to which the limit can be taken 
and the theories exhibit better convergence behavior for all energy regions.
Again, the time function has a non-trivial dependence on the parameter $\l$. What is 
perhaps surprising is that such function, constructed entirely within the Hamiltonian formalism, corresponds precisely 
to the parameter one calls cosmic {\it proper time} in the space-time description and has a direct physical 
interpretation. With this choice, convergence is uniform and the limit exists. One does not, however, recover the 
classical dynamics. Instead, the limit corresponds to a classical contracting solution connected  
(in singular, but continuous way) to another classical expanding solution.

As we mentioned in the Introduction, the question we have posed is the classical equivalent of considering the 
continuum limit for constrained quantum polymeric theories. Such program is yet to be established. Perhaps the most 
important lesson we can draw from our analysis is that one needs to choose very carefully the observables --in this 
case, a time observable-- in order to be able to consider the question of convergence. In the quantum theory, the 
corresponding choice will correspond to appropriate operators whose expectation values will exhibit the desired 
convergence properties. We can only hope that the lessons learnt here will be helpful in such endeavor.

\section*{Acknowledgments}
%\vskip0.5cm
\noindent
We thank J.A. Zapata for discussions and comments during the early part of this project. 
This work was in part supported by DGAPA-UNAM IN103610 grant, by NSF
PHY0854743, the Eberly Research Funds of Penn State and by CIC, UMSNH.

\end{document}